# Dependence of the Radical Dynamics on the Beam Temporal Profile in FLASH Radiotherapy

Jianhan Sun [1,2], Xianghui Kong [3], Jianfeng Lv [1,4], Xiaodong Liu [5], Jinghui Wang [4,6], Chen Lin [1,4], Tian Li [3], Yibao Zhang [2*], Senlin Huang [1*]

[1]*State Key Laboratory of Nuclear Physics and Technology and Institute of Heavy Ion Physics, School of Physics, Peking University, Beijing 100871, China*

[2]*Key Laboratory of Carcinogenesis and Translational Research (Ministry of Education), Department of Radiation Oncology, Peking University Cancer Hospital & Institute, Beijing 100142, China*

[3]*Department of Health Technology and Informatics, The Hong Kong Polytechnic University, Hong Kong 100872, China*

[4]*Beijing Laser Acceleration Innovation Center, Beijing 101407, China*

[5]*State Key Laboratory of Heavy Oil Processing, China University of Petroleum, Qingdao, Shandong 266580, China*

[6]*Guangdong Institute of Laser Plasma Accelerator Technology, 510540, China*

a) Authors to whom correspondence should be addressed: zhangyibao@pku.edu.cn (Yibao Zhang), huangsl@pku.edu.cn (Senlin Huang)

b) Jianhan Sun and Xianghui Kong contributed equally to this work.

# Abstract

**Purpose:** This study aims to investigate the impact of the beam temporal profile on the radical dynamics and inter-track interactions of FLASH radiotherapy, supporting parameter optimization for the equipment development, radio-biological experiments and clinical implementation.

**Methods and Materials:** Monte-Carlo simulations based on the independent reaction time (IRT) method were performed to analyze the dynamics after irradiation, including single-pulse or multi-pulses irradiation, pulse repetition rate, pulse width and dose. The physicochemical experiments were performed to measure the hydrated electron lifetimes for validation. The generation and recombination of hydroxyl radicals and hydrated electrons were recorded under 6 MeV electron irradiation with varying beam temporal profiles. The radial distributions of the radicals were statistically analyzed, and the corresponding dose-averaged linear energy transfer ($LET_d$) and track-averaged linear energy transfer ($LET_t$) were calculated. The inter-track interactions were assessed through a mathematical model.

**Results:** The spatial distribution and temporal evolution of radicals were significantly affected by the beam time profiles. Compared with multi-pulses irradiation, single-pulse irradiation mode with a pulse width less than 1/10 of the radical lifetime, a repetition interval longer than the radical lifetime, and a dose exceeding 1 Gy/pulse can lead to rapid consumption of radicals within the first 30% of their lifetime, hence reduced the residual radical content. Instantaneous high dose rates induced overlapping of radical tracks. When the single-pulse dose exceeded 1 Gy, the overlap probability approached 100%, aligning with the dose threshold for the instantaneous radical combination.

**Conclusion:** Under a low-duty cycle and high instantaneous dose-rate time profile, the radicals were rapidly consumed through track overlap hence reduced damage to normal tissues, inducing FLASH effect. The optimized time profile can be used to guide the development of equipment and parameter settings in clinical practice to maximize the FLASH effect, such as the laser accelerators and superconducting photocathode guns.

## 1. Introduction

FLASH radiotherapy (FLASH-RT) is a novel treatment approach characterized by ultrahigh dose rates delivered within extremely short time intervals[1]. FLASH-RT can reduce normal tissue toxicity, maintain the therapeutic efficacy, and improve the treatment efficiency[2], as observed in the preliminary clinical trials[3, 4]. Several hypotheses have been proposed to explain the mechanism of FLASH-RT, including depletion of oxygen in tissues[5, 6], accelerated combination of radicals[7–10,] preservation of DNA integrity in normal cells[11–13], and the mitochondria-related hypothesis[14, 15], yet no consensus has been reached. The characterization of FLASH involved several interdependent physical beams of light parameters. Recent analyses suggested that in addition to the average dose rate ($\geq$40 Gy/s), the instantaneous dose rate and total duration of exposure may also affect the FLASH effect[16].

Various clinical or experimental systems hav e been proposed to deliver FLASH-RT, where the beam time profiles varied dramatically, as shown in Figure 1(a)[17]. For instance, the modified medical linear accelerators achieved an average dose rate of more than 200 Gy/s, a pulse width of several microseconds (μs) and a single-pulse dose of approximately 1 Gy[18–20]. The synchrotrons can generate proton or heavy ion pulses of approximate order of picoseconds (ps). The typical radio frequency of the third-generation synchrotron radiation sources is in the range of hundreds of megahertz (MHz) range[21, 22]. The first carbon ion-based FLASH-RT using a synchrotron[23] achieved an average dose rate of 70 Gy/s [24]. The superconducting accelerators can deliver ultrashort (ps class), high repetition rates (MHz--GHz), and high beam current[25, 26]. Superconducting accelerators can operate in continuous wave mode, enabling flexible beam delivery tailored to clinical needs. The first X-ray FLASH effect experiment was conducted on the Chengdu THz Free Electron Laser facility (CTFEL)[27]. In addition, laser accelerators utilize ultra-intense laser pulses to generate femtosecond (fs) pulse beams to achieve the ultrahigh dose rates for FLASH-RT[28]. Although the repetition rate is only several hertz (Hz), they can deposit almost the entire radiation dose (> 20 Gy) in a single shot, resulting in an instantaneous dose rate as high as $10^9$ Gy/s[29, 30].

The quantitative impact of various beam time profiles on the FLASH effect remains unclear[31]. Previous studies primarily focused on the millisecond (ms)-scale time fractionation[32], leaving finer timescales of ps to μs unexplored. However, these finer temporal profiles cover the entire chemical stage, which is particularly relevant to the dynamics of radical generation, combination, and diffusion, as shown in figure 1(b). The lack of detailed experimental and simulation evidences at these time scales have limited the understanding of the underlying mechanisms, hence impeded the efficient equipment development and effective clinical application of FLASH-RT.

By modeling the energy deposition, radical dynamics, and spatiotemporal distributions, Monte-Carlo (MC) methods have been extensively employed to simulate the particle–matter interactions and the subsequent biological effects of radiation in the FLASH effect [32–35]. Using ultra-high dose rates, the delivery time of particles can be reduced to the μs scale, which is comparable to the lifetime of chemical species such as reactive oxygen species produced by the reaction[34]. TOPAS-nBio further enabled time-dependent simulations across a wide temporal range from ps to ms[35], using which Ramos-Méndez et al. revealed LET-dependent reactant yields under proton irradiation at ultrahigh dose rates and evaluated the effects of inter-track interactions[35]. Shannon J Thompson et al. provided mathematical models for calculating inter-track interactions and their correlations with the spatial separation between incident protons[34].

To optimize the dose–time distributions and investigate the underlying mechanisms of FLASH effect, this MC-based study systematically analyzed the impact of the electron beam temporal profile on the FLASH effect, by calculating the radical production, combination, and energy deposition induced by various pulse width, repetition rate, and pulse dose.

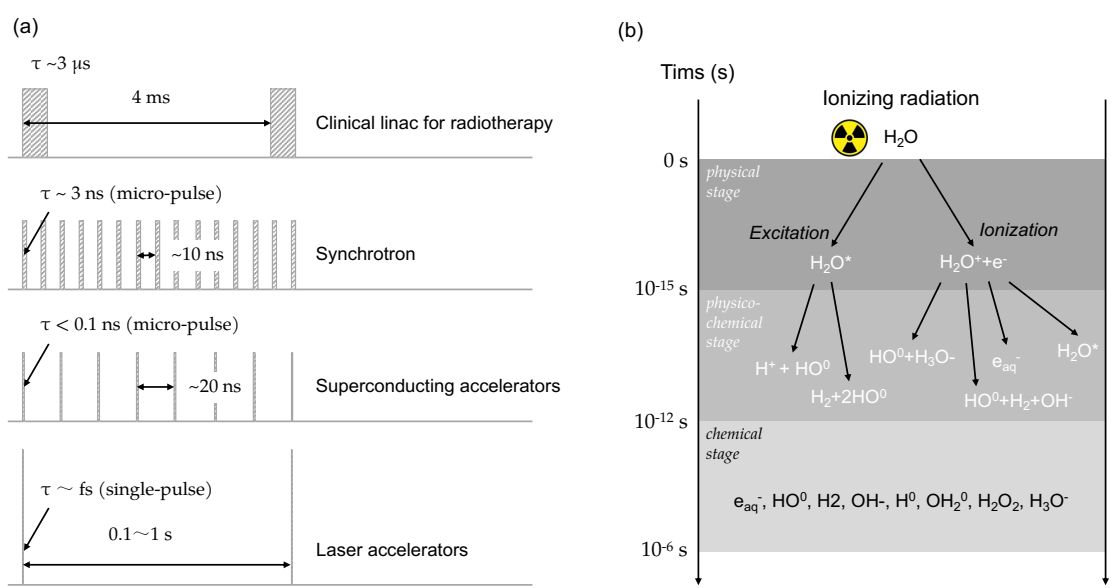

Figure 1. (a) Beam time profiles of different accelerators that have be used for FLASH radiotherapy. (b) The time scale corresponding to the water radiolysis process.

## 2. Methods and Materials

### 2.1 Simulation setup

Using the OpenTOPAS v4.0.0 toolkit (Faddegon B et al 2020[1]) and the TOPAS-nBio v3.0 extension (Schuemann et al 2019[2]), a MC model was developed to simulate the electron irradiation of ultra-high dose rate and subsequent radical diffusion in water. The technical simulation details are provided in the Supplementary Materials Table S1, including the packages and chemical reaction constants. The incident electron trajectory and generated radical distribution are shown in figure S1(a), and the total model is shown in figure S1(b). To reduce the contingency of the results, the simulations were repeated using 10 random seeds over 1 Gy, and 30 random seeds were used for other low-dose exposures.

2.2 Track overlap model

A simplified track overlap model was built to investigate the potential interactions of electron intertrack across a range of doses and chemical stage timepoints[34]. This model was used to predict the geometric probability of the track overlap as an indicator of potential physical or chemical inter-track interactions. The radial distance from the central axis of the beam was calculated for the energy deposition position and radical generation position. The radial distance encompassing 80% of the radicals was defined as the track radius of the chemical substance. These parameters were used to determine the evolution of the track in relation to the energy and diffusion times. This model assumed that all tracks arrived simultaneously and immediately, which determined whether the physical or chemical tracks were spatially close enough to interact hence provided an upper limit for interactions between tracks. The statistical likelihood of inter-track overlap probability of $N$ tracks can be calculated from:

$$P(N, r, R) = 1 - \left(1 - \frac{\pi r^2}{\pi R^2}\right)^N \tag{1}$$

where $N$ denotes the quantity of particle trajectories, $r$ represents the radius of either particle or free radical tracks, and $R$ specifies the radius of the target area. Alternatively, the expected numbers of overlaps per individual track can be computed using the follow equation:

$$Expected\ Number\ of\ Overlaps = \frac{\pi r^2}{\pi R^2} \frac{(N-1)}{2} \tag{2}$$

2.3 LET calculation based on Monte-Carlo methods

The dose-averaged linear energy transfer ($LET_d$) and track-averaged linear energy transfer ($LET_t$) were calculated as weighted averages of the LET in the target area according to the dose. Considering the interactions between the small sensitive biomolecules (such as DNA) and the radiation particles often occur in local regions of the particle track, $LET_t$ can well describe the average energy transfer of individual particles at the microscopic level, as well as how radiation directly interacts with biomolecules. $LET_t$ and $LET_d$ were calculated using the following formulas[38]:

$$LET_t = \sum_{i=1}^{n}\left(\frac{\varepsilon_i}{l_i}\right)w_{i,t} = \frac{\sum_{i=1}^{n}\left(\frac{\varepsilon_i}{l_i}\right)l_i}{\sum_{i=1}^{n}l_i} = \frac{\sum_{i=1}^{n}\varepsilon_i}{\sum_{i=1}^{n}l_i} \tag{3}$$

$$LET_d = \sum_{i=1}^{n} \left(\frac{\varepsilon_i}{l_i}\right) w_{i,d} = \frac{\sum_{i=1}^{n} \left(\frac{\varepsilon_i}{l_i}\right) \varepsilon_i}{\sum_{i=1}^{n} \varepsilon_i} = \frac{\sum_{i=1}^{n} \frac{\varepsilon_i^2}{l_i}}{\sum_{i=1}^{n} \varepsilon_i} \qquad (4)$$

where $\varepsilon_i$ denotes the energy deposition of the $i^{th}$ charged particle with tracking step length $l_i$; $n$ is the total number of charged particles; $w_{i,t}$ and $w_{i,d}$ are the tracking length weighting factor and the dose weighting factors of the $i^{th}$ event.

2.4 Lifetime detection of hydrated electrons ($e_{aq}^-$)

A 632 nm, 100 mW helium–neon laser was used to measure the lifetime of $e_{aq}^-$, which absorption spectrum spanned from 500 to 800 nm, peaking at 715 nm. When $e_{aq}^-$ was generated along the laser path, the transmitted laser intensity decreased. As $e_{aq}^-$ decays rapidly owing to the chemical reactions, the transmitted laser intensity recovers. The $e_{aq}^-$ lifetime was determined by recording the intensity change over time. As shown in Figure 2(a), the experiment was conducted in a 5x5x5 cm³ cubic glass cell filled with 100 mL water. The oxygen was removed by bubbling Nitrogen for 30 min, and the cells were sealed. The horizontally incident laser was 1 cm below the water surface. The pulse width of the 6 MeV electron beam was approximately 3 μs, and a 10% attenuation filter was used. The detector was 4 m away, which recorded the intensity via an oscilloscope, avoiding Cherenkov radiation interference.

As shown in figure 2(b), the Monte Carlo simulation suggested that the oxygen had a rapid scavenging effect on $e_{aq}^-$, hence the lifetime of $e_{aq}^-$ depended on the oxygen concentration, which was governed by a second-order rate constant ($k = 1.9 \times 10^{10}$ M⁻¹s⁻¹). Therefore, to control the oxygen content in the system, the dissolved oxygen concentration in the aqueous samples was maintained at 1.7 mg/L ($5.31 \times 10^{-5}$ M), corresponding to the physiological oxygen partial pressures (pO₂) of 10–60 mmHg ($1.3 \times 10^{-5}$ M to $7.8 \times 10^{-5}$ M at 37°C) [39, 40]. This oxygen concentration was also selected for subsequent simulations to investigate the radical dynamics under physiological conditions.

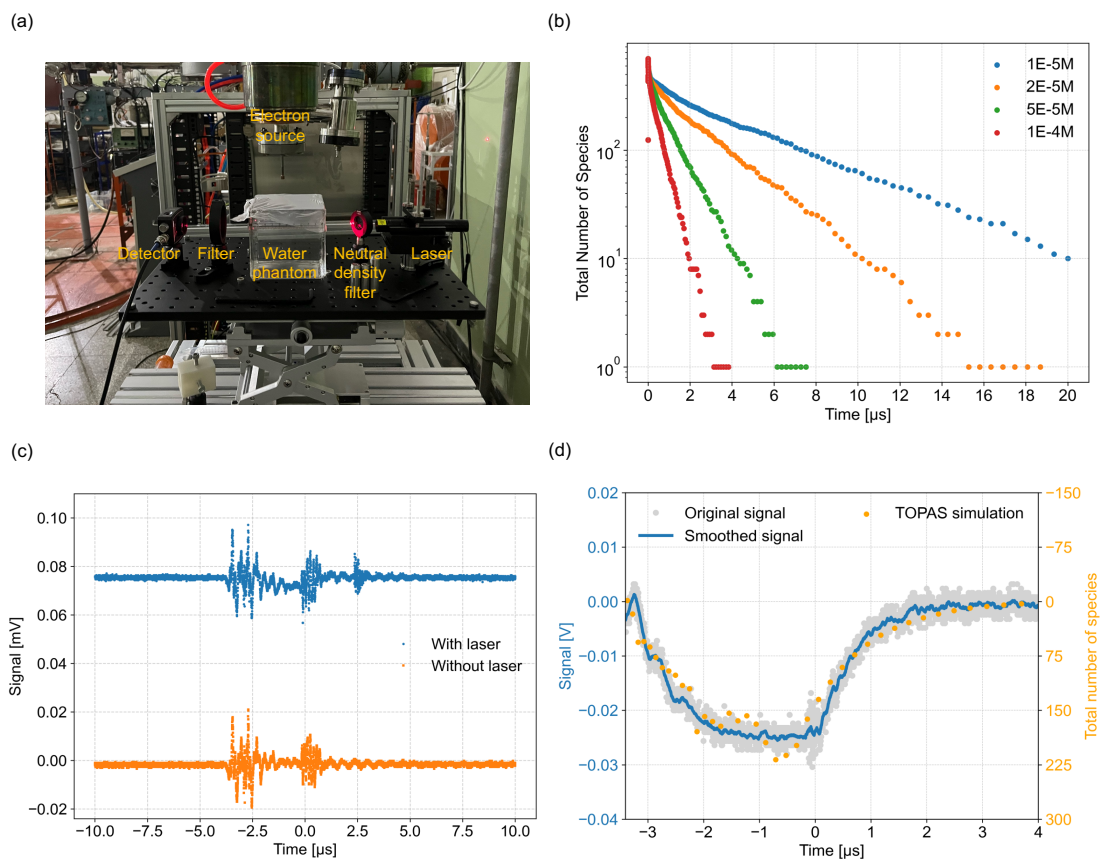

Figure 2. Experiment settings of $e_{aq}^-$ lifetime detection. (a) Schematic of the experimental apparatus. (b) Relationship between MC-simulated oxygen content and the lifetime of $e_{aq}^-$ radicals. The blue, orange, green and red dotted-lines represent the oxygen content in water of $1\times10^{-5}$ Mol · L$^{-1}$, $2\times10^{-5}$ Mol · L$^{-1}$, $5\times10^{-5}$ Mol · L$^{-1}$ and $1\times10^{-4}$ Mol · L$^{-1}$ respectively. (c) Comparison of the measured signals with and without laser excitation. (d) Comparison between the experimental results and simulated results. The blue line represents the time-resolved signal waveform after background subtraction, indicating the produce and decay profile of $e_{aq}^-$. The yellow points represent the TOPAS-simulated number of $e_{aq}^-$ over time.

## 3. Results

### 3.1 Detection of the $e_{aq}^-$ lifetime

To validate the MC-simulated results, the lifetime of $e_{aq}^-$ in water under electron beam irradiation was experimentally measured. Figure 2(c) shows the detected raw signals with and without the laser trigger. Their differences reflect the changes of $e_{aq}^-$ radicals before and after irradiation, which are shown as the gray dots (the original values) and the blue curve (after smooth filtering) in figure 2 (d).

## 3.2 Generation and evolution of radicals under different temporal profiles

Various temporal characteristics of pulsed beam delivery were investigated, including different irradiation mode (single-pulse versus multi-pulse), pulse repetition frequency in multi-pulses operation, individual pulse duration, and cumulative radiation dose. The temporal evolutions of two typical radicals, i.e., hydroxyl radicals (·OH) and hydrated electrons ($e_{aq}^-$), are depicted in Figure 3. Figure 3(a) and (b) illustrate the distinctly different evolution trends of these two radicals under single-pulse and repetitive-pulse irradiation using a total dose of 10 Gy. Their lifetimes were approximately 10 ms and 10 μs, respectively. The repetition rates of multi-pulses were determined by 1/10, 1/100, and 1/1000 of the lifetime of this radical, i.e., 1 kHz to 100 kHz for ·OH and 1 MHz to 100 MHz for $e_{aq}^-$. Under single-pulse irradiation, the radicals were generated rapidly and decayed within a few ms (·OH) or μs ($e_{aq}^-$), approximately 30% of their lifetime. In contrast, multi-pulses irradiation resulted in a stepwise increase in the radical concentration over time. Different repetition rates in the multi-pulses mode had no obvious effect on the results, denying its role as a dominant factor when the repetition cycle of pulses was less than 1/10 of the lifetime of radicals.

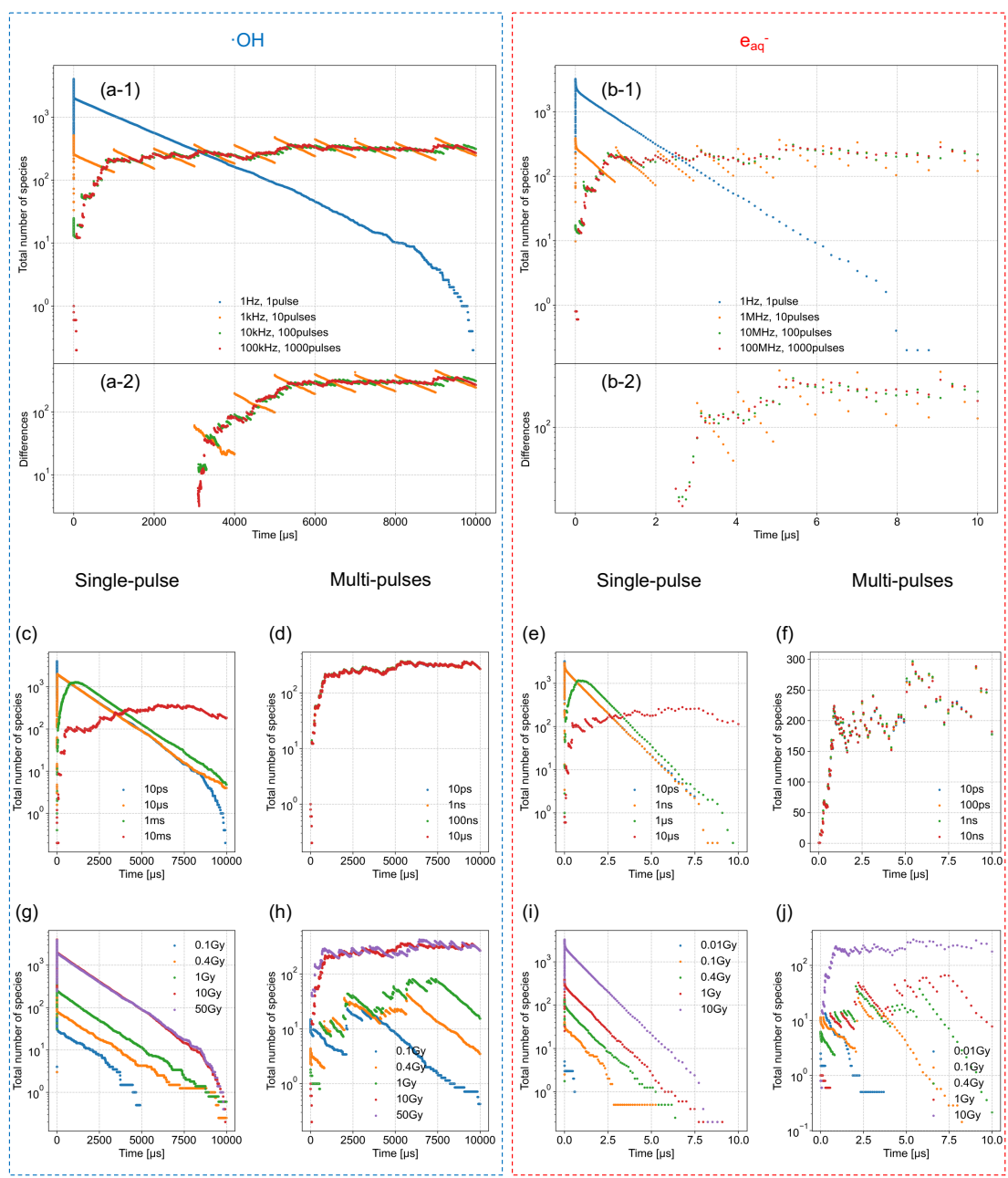

Figure 3. Impact of different temporal profiles on the production of radicals: Variations in ·OH (a-1) and $e_{aq}^-$ (b-1) under single-pulse and multi-pulses irradiation with different repetition frequencies, using pulse width of 10 ps and total dose of 10 Gy; (a-2, b-2) the difference between each line and the single-pulse cases shown as blue lines; Variations in ·OH (c, d) and $e_{aq}^-$ (e, f) under single-pulse and multi-pulses irradiation of 10 Gy with different pulse widths; Variations in ·OH (g, h) and $e_{aq}^-$ (i, j) under single-pulse and multi-pulses irradiation with different total doses, using pulse width of 10 ps. (i, j).

The effects of the other time profile parameters are shown in figure 3 (c)–(j). Figure 3 (c) and (e) illustrate the impact of the micropulse pulse width on the single-pulse irradiation. For pulses with widths less than 1/100 of the radical lifetime (e.g., 10 ps or 10 μs for ·OH; 10 ps or 1 ns for $e_{aq}^-$), the radical generation and decay curves were nearly identical, reflecting instantaneous radical generation followed by rapid combination and consumption. However, for a pulse width extending to 1/10 of the radical lifetime, such as 1 ms for ·OH and 1 μs for $e_{aq}^-$, the radical production became gradual, which delayed the consumption and left more residue, indicating the waning advantage of single-pulse irradiation. In contrast, figure 3 (d) and (f) reveal minimal effects of variations in pulse width, denying the significant role of the temporal profile of individual micro-pulses during the multi-pulse irradiation. The total irradiation dose also affected the evolution of the two radicals, as shown in figure 2 (g)–(j).

Additionally, variations in the electron beam energies from 1 MeV to 100 MeV had little effect on the total radical yields and their temporal evolutions, as depicted in figure S1 in the supplementary materials. This energy-independent behavior can be explained by the near-constant LET values of electrons in this range, clearly indicating that beam energy is not a primary determinant of radical generation in FLASH radiotherapy.

3.3 Spatial distribution of energy deposition and chemical species

To further investigate the differences in radical generation and evolution under different temporal profiles, the spatiotemporal distributions of energy deposition and generated radicals under a single-pulse were simulated and analyzed, as shown in figure 4. All incident particles were assumed to be produced instantaneously, corresponding to an infinite instantaneous dose rate. As an analysis of the radial distribution of energy deposition, figure 4(a) demonstrates overlapping energy deposition profiles at different time points after pulse delivery. Using 10, 100, 1000 and 10000 particles, the $LET_d$ (keV/μm) was calculated as 0.120, 0.138, 0.144, and 0.143 respectively, and the $LET_t$ (keV/μm) was calculated as 0.102, 0.096, 0.098, and 0.097 respectively, based on the methods described in Section 2.3.

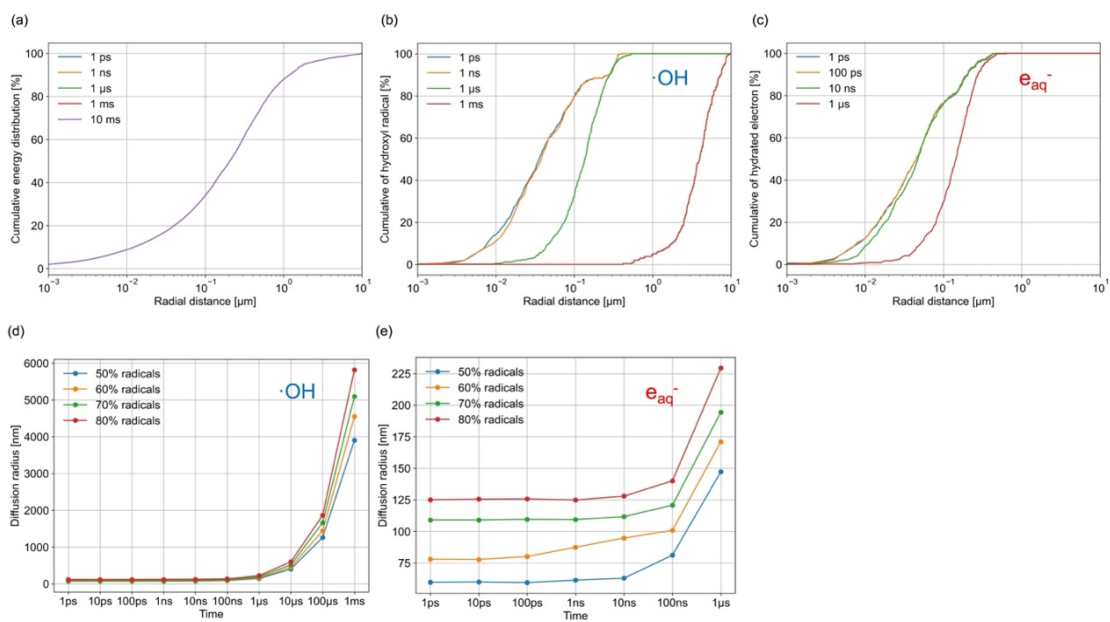

Figure 4. Radial distributions of energy deposition and chemical species ($\cdot$OH and $e_{aq}^-$) distributions. (a) Cumulative energy deposition at different time points after pulse delivery for pulse widths ranging from 1 ps to 1 ms. (b, c) Radial cumulative distributions of $\cdot$OH and $e_{aq}^-$ around an electron track at different time points. (d, e) Diffusion radii defined by different proportions of $\cdot$OH and $e_{aq}^-$ in the envelope, and the moment at 1 μs was taken.

To understand the radical production dynamics, the spatial distributions of $\cdot$OH and $e_{aq}^-$ radicals were analyzed. Figure 4 (d) and (e) depict the radial cumulative distributions of $\cdot$OH and $e_{aq}^-$ around an electron track at different time points. Figure 4(f) and (g) show the radial distance of the number of radicals within a certain proportion of the envelope around the incident track at different times. The ratios of the envelope radii of different proportions remained constant throughout the monitoring process. For $e_{aq}^-$ as an example, the ratio of the 80% envelope radius to the 50% envelope radius was 1.91 at 1 ps and 1.93 at 1 μs. These results indicated that the radicals diffused outward at the same rate, and the radicals adjacent to the center did not undergo self-combination or accelerated consumption because of their transiently high concentration. Considering the negligible impact of time profile below 1 μs on the evolution of the two radicals as shown in Section 3.2, the radial distribution was selected at 1 μs after irradiation and the orbital radius of the radicals was defined by the 80% envelope range.

3.4 Overlap probability and expected track overlap of chemical species

The total irradiation doses were calculated for different numbers of incident particles using MC simulations. The dose on the horizontal axis was used to represent the particle density. Using the model described in Section 2.2, the probability and expected number of $\cdot$OH and $e_{aq}^-$ chemical track overlaps were determined, as shown in figure 5. At low doses (e.g., $10^{-3}$ Gy), the overlap probability was near zero. As the dose increased, the probability rose sharply, surpassing 50% at approximately $10^{-1}$ Gy, and approaching 100% at 1 Gy. The expected number of overlaps followed a similar pattern, remaining negligible at low doses but increasing rapidly beyond $10^{-2}$ Gy, reaching 1 at 0.4 Gy, and reaching 2–3 above 1 Gy.

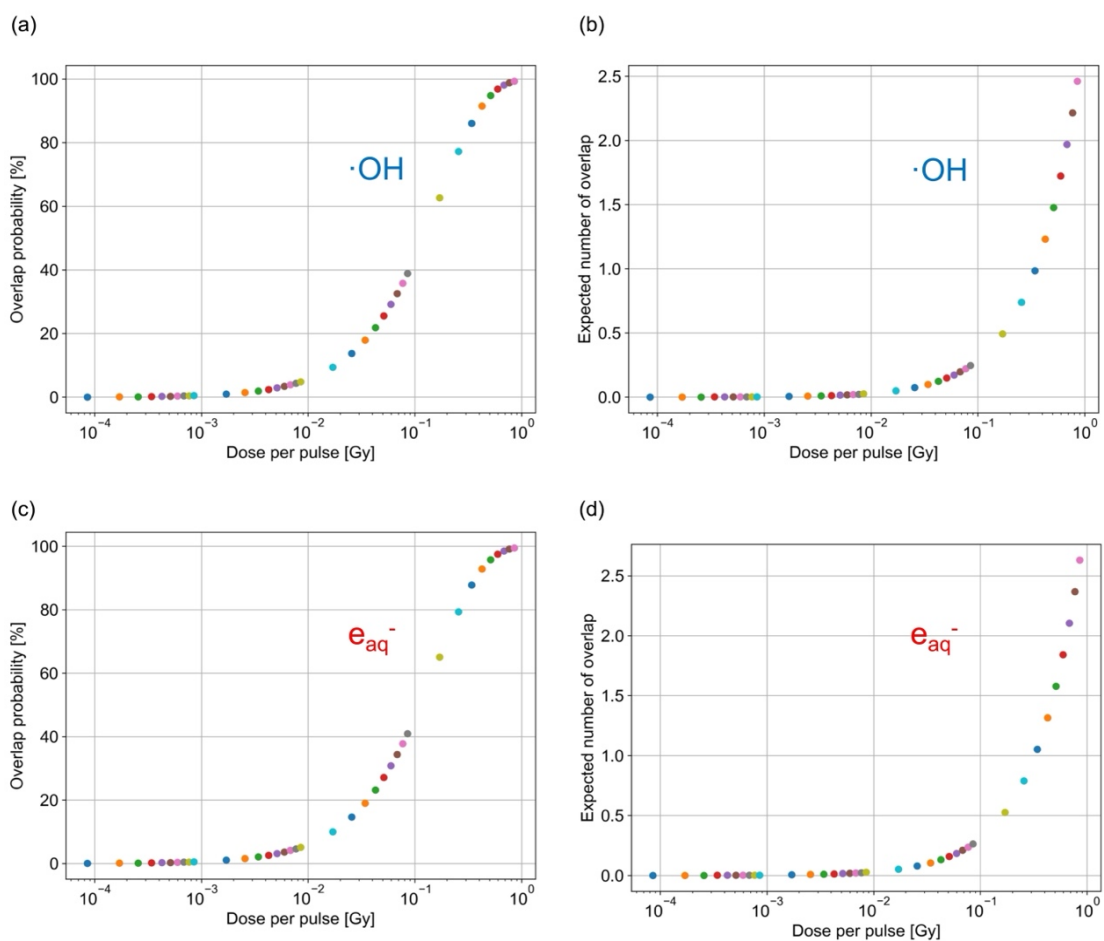

Figure 5. The calculated overlap probabilities and the expected number of overlaps per track for ·OH (a, b) and $e_{aq}^-$ (c, d) as a function of dose per pulse.

## 4. Discussion

By analyzing the mechanisms of inter-track interactions based on MC simulations, this study investigated the potential impact of electron beam temporal profiles ranging from ps to ms timescale on the FLASH effect. The accuracy of the MC mode was validated by the consistency with the measured radical evolution process as shown in figure 2(d). The investigated parameters of the beam temporal profile included the repetition rate, micropulse width, and bunch particle number (or dose).

The differences in the radical generation and the combination of single-pulse and multi-pulses irradiation have significant implications for FLASH radiotherapy[1, 32]. The underlying mechanisms can be attributed to the rapid combination and faster consumption of radicals as a result of higher instantaneous radical concentrations induced by the single-pulse irradiation. For example, the reaction rate of $e_{aq}^-$ with oxygen ($k = 1.9 \times 10^{10}$ M$^{-1}$s$^{-1}$) dominates under single-pulse conditions, leading to rapid depletion, as shown in figure 3(b). In contrast, multi-pulses irradiation allows for radical replenishment, reducing scavenging dominance and prolonging the activity of species such as ·OH and $e_{aq}^-$. This staggered energy delivery reduced the impact of rapid scavenging, allowing the system to reach a quasi-equilibrium state, where balanced radical decay and regeneration occurred. Compared with multi-pulses irradiation, this rapid radical combination may reduce the level of DNA damage in normal cells. In contrast, cancer cells typically have 2–4 times higher levels of active metal ions, particularly $Fe^{2+}$, which enhances Fenton reactions and amplifies oxidative damage[41, 42]. Coupled with weaker antioxidant systems, this results in poor ROS scavenging[2, 8, 43]. Therefore, the toxicity of radiation to cancer cells was preserved, yet the normal tissues were better preserved. This finding is consistent with "the peroxyl radical recombination-antioxidants hypothesis" for FLASH effect[44]. This finding is also supported by the recent studies which reported that rescanning proton irradiation increased the toxicity compared with single irradiation at the same field dose rate, shifting the dose rate–response curve to the right[45].

The radical combination-induced difference may be affected by the temporal profile and the total dose of irradiation. Because the lifetimes of the two radicals, ·OH and $e_{aq}^-$, differ by three orders of magnitude, their matching timescales are quite different. However, their evolutionary patterns are similar. By comparing figure 3(a) to (j), it was observed that the rapid consumption of the radicals requires the simultaneous fulfillment of the following conditions: (I) The repetition interval of the pulse beam is greater than the lifetime of the radicals; (II) The pulse beam width is less than 1/10 of the radical lifetime scale, i.e., the beam duty cycle should be less than 10%; (III) The dose of one single-pulse is greater than 1 Gy. Under these conditions, the temporal structure within a single-pulse does not significantly affect the variation in radicals throughout their lifetime. These results may explain some previous studies, where the FLASH effect was not observed. The negative results may be attributed to the beam time profiles, failing to meet the requirements for instantaneous recombination of radicals. For instance, Kim *et al*. reported similar cell damage to Lewis lung cancer cells in mice and pancreatic endothelial cells in SVR mice using pulse interval of 5 ms[46], which could not fully consume the radicals but accumulated continuously.

The minimum single-pulse dose threshold was quantitatively validated through computational modeling of inter-track overlap dynamics. This work demonstrated that significant track overlap was necessary to achieve a critical concentration of radical conversion. As shown in figure 5, when the single-pulse dose reaches 0.4 Gy, the expected number of overlaps per track reached 1. When the dose per pulse exceeded 1 Gy, the track overlap probability began to approach 100%. This result aligned well with the threshold for different radical combination: when the total dose was greater than 0.4 Gy, the lifetime of the radicals under a single-pulse irradiation began to converge. When it was greater than 1 Gy, the continuous accumulation of radicals under multi-pulses irradiation cannot be consumed rapidly. Thus, radical track overlap represents a key contributor to the increased radical recombination frequency. Because the lifetime of radicals can last up to μs or even longer, it is possible to activate chemical-track overlap and affect the lifetime and distribution of radicals. If roughly estimated by the rapid combination of ·OH, the minimum average dose rate required for FLASH effect was calculated as 0.4 Gy per 10 ms in this work, i.e., 40 Gy/s, which is exactly consistent with the well-known standard for FLASH-RT[1].

As a potential technical guidance for developing FLASH equipment and setting clinical parameters based on the findings of this study, figure S2 in Supplementary Materials illustrates two possible beam generation schemes for future FLASH-RT. The irradiation beams with single-pulse profiles, short duty cycles, and high instantaneous currents are favorable for the instantaneous combination and consumption of radicals. However, the characteristic lifetimes of ·OH (microsecond) and $e_{aq}^-$ (nanosecond) differ by more than three orders of magnitude, creating a fundamental temporal mismatch that presents substantial technical challenges for achieving synchronous radical recombination dynamics. Therefore, laser accelerators (figure S2(a)) can be used to produce a single shot of radiation with an extremely high instantaneous dose and a sub-ps width, enabling the rapid combination of both radicals within a single-pulse[47]. Alternatively, a superconducting photocathode gun (figure S2(b)) can be used to generate flexible and adjustable macro-micro pulse-time profiles[25, 48]. For instance, the electron beam used in this study was produced via the photoelectric effect based on the superconducting photocathode gun, forming micropulses with pulse widths of ps. Based on the dual acousto-optic modulators (AOMs) for pulse selection, the micro-pulse envelope can form pulse trains with adjustable repetition rates and duty cycles, known as "macro-pulses" [49]. Therefore, by controlling the time distribution of the driving laser, the micropulses and macropulses can be adjusted to meet the combination conditions of $e_{aq}^-$ and ·OH, respectively, hence achieve a rapid combination of both radicals simultaneously. Typical photocathode guns such as DC-SRF-II can achieve this beam time profile, which has been successfully used in the cell experiments and preclinical research for FLASH-RT[14, 15].

## 5. Conclusion

This study systematically analyzed the impact of the beam temporal profile on FLASH radiotherapy via MC simulations. Compared with multi-pulses irradiation, single-pulse irradiation rapidly scavenged radicals and left fewer residual species. This phenomenon was ascribable to the overlap of radical tracks, leading to instantaneous recombination. This rapid recombination could be one of the underlying mechanisms reducing damage to the normal tissues in FLASH-RT. Three conditions must be met simultaneously to achieve this rapid recombination: short pulses, long repetition intervals, and high instantaneous current. Through the generation of specialized time profiles, both laser acceleration and superconducting photocathode guns can facilitate the simultaneous combination of radicals with different lifetimes, such as ·OH and $e_{aq}^-$. These findings underscore the significance of beam temporal profiles in the FLASH effect, offering potential guidance for developing advanced FLASH delivery systems and optimizing treatment protocols in the clinical practice.